\documentstyle[ltwol,epsfig,axodraw,graphicx,psfrag]{article}

\def\be{\begin{equation}}
\def\ee{\end{equation}}
\def\bea{\begin{eqnarray}}
\def\eea{\end{eqnarray}}

\begin{document}

\title{Effect of Supersymmetric phases
on the Direct CP Asymmetry of $B\to X_d\gamma$}

\author{A.G.\ Akeroyd, S.\ Recksiegel}
\address{Theory Group, KEK, Tsukuba, Ibaraki 305-0801, Japan\\
E-mail: akeroyd@post.kek.jp, stefan@post.kek.jp}

\author{Y.-Y.\ Keum}
\address{Institute of Physics, Academia Sinica,
Nankang, Taipei, Taiwan , 11529, R.O.C.\\E-mail: keum@phys.sinica.edu.tw}


\twocolumn[\maketitle\abstracts{We investigate 
the effect of supersymmetric CP violating phases on 
the inclusive decay $B\to X_d\gamma$. Although such a decay
contains a large background from $B\to X_s\gamma$, if isolated 
it may exhibit sizeable CP violation, 
both in the Standard Model (SM) and in the context of models beyond
the SM. With unconstrained supersymmetric CP violating phases
we show that the direct CP asymmetry 
(${\cal A}_{CP}$) lies in the region $-40\% \le {\cal A}_{CP}\le 
40\%$, where a positive asymmetry would constitute a clear 
signal of physics beyond the SM. Even if a direct 
measurement of $B\to X_d\gamma$ proves too difficult experimentally,
its asymmetry contributes non-negligibly to the measurements 
of ${\cal A}_{CP}$ for $B\to X_s\gamma$, and thus should be included
in future analyses. We show that there may be both constructive and
destructive interference between  ${\cal A}^{d\gamma}_{CP}$ and
${\cal A}^{s\gamma}_{CP}$.}]

\section{Introduction}

Theoretical studies of rare decays of $b$ quarks have attracted
increasing attention with the recent turn-on of the $B$ factories at
KEK and SLAC. In this talk we are concerned 
with the rare decay $b\to d\gamma$ which proceeds via an electromagnetic
penguin diagram, and is sensitive to the CKM matrix element 
$V_{td}$. The work on which this talk is based has been published
in \cite{AkeKeRe}.

The current measurement of ${\cal A}_{CP}$
for $b\to s\gamma$ by the CLEO Collaboration \cite{CLEO} is
sensitive to events from $b\to d\gamma$. Therefore knowledge of
${\cal A}_{CP}$ for $b\to d\gamma$ is essential, in order to
compare experimental data with the theoretical prediction 
in a given model.

We are interested in the effect of unconstrained supersymmetric (SUSY)
CP violating phases on the inclusive decay BR$(B\to X_d\gamma)$. 
We will be working in the context of the effective SUSY model proposed 
in \cite{effSUSY}. Such a model allows one to consider the full 
impact of the phases on the rare decays of $B$ mesons, while simultaneously
satisfying the stringent bounds on the Electric Dipole Moments 
of the electron and neutron.

\boldmath
\section{The decays $b\to d\gamma$ and $b\to s\gamma$}
\unboldmath

Much theoretical study has been devoted to the decay $b\to s\gamma$ 
due to its sensitivity to physics beyond the SM
\cite{bsytheory}. Exclusive channels ($B\to K^*\gamma$ etc.) 
and the inclusive channel have been measured at CLEO, ALEPH, BELLE and 
BaBar \cite{bsymeasure}.
The related decay $b\to d\gamma$ has received less attention
although is expected to be observed at the $B$ factories, at least
in some exclusive channels. 

Ref. \cite{plb429} calculated BR$(B\to X_d\gamma$)
in the context of the SM. It was shown that the ratio $R$ defined by
\begin{equation}
R={BR(B\to X_d\gamma)\over {BR(B\to X_s\gamma)}}
\end{equation}
is expected to be in the range $0.017 < R < 0.074$, corresponding
to BR$(B\to X_d\gamma)$ of order $10^{-5}\to 10^{-6}$. 
With $10^8$ $B\overline B$ pairs expected
from the $B$ factories, one would be able to produce $10^2\to 10^3$
$b\to d\gamma$ transitions.  In the ratio $R$ most of the theoretical
uncertainties cancel, and hence $R$ may provide a theoretically clean
way of extracting the ratio $|V_{td}/V_{ts}|$.

The CP asymmetry (${\cal A}_{CP}$), defined by\footnote{
Note that our definition ${\cal A}_{CP}$ contains a relative
minus sign compared to that used in \cite{plb429},\cite{plb460},
\cite{CLEO}}
\begin{equation}
{\cal A}^{d\gamma(s\gamma)}_{CP}={{\Gamma(\overline B\to X_{d(s)}\gamma)-
\Gamma(B\to X_{\overline d(\overline s)}\gamma)}
\over {\Gamma(\overline B\to X_{d(s)}\gamma)+\Gamma(B\to X_{\overline
d(\overline s)}\gamma)}}={\Delta\Gamma_{d(s)}\over \Gamma^{tot}_{d(s)}}
\label{ACPdef} \end{equation}
is expected to lie in the range $-7\%\le {\cal A}_{CP}^{d\gamma}\le -35\%$ 
in the SM \cite{plb429}, where the uncertainty arises from 
varying $\rho$ and $\eta$
in their allowed regions. Also included is the scale dependence $(\mu_b)$ 
of ${\cal A}_{CP}^{d\gamma}$ which occurs from varying
$m_b/2\le \mu_b \le 2m_b$. For definiteness we fix
$\mu_b=4.8$ GeV, and find $-5\%\le A_{CP}^{d\gamma}\le -28\%$.
Therefore  ${\cal A}_{CP}^{d\gamma}$ is much larger than
${\cal A}_{CP}^{s\gamma}$ ($\le 0.6\%$). By estimating values for detection 
efficiencies, it has been argued in \cite{KSW}
that ${\cal A}^{d\gamma}_{CP}$ may be statistically more accessible than
${\cal A}^{s\gamma}_{CP}$, at least in the context of the SM.
This analysis assumes that 
$B\to X_d\gamma$ can be clearly isolated from $B\to X_s\gamma$.

However, it is known that isolating the signal $B\to X_d\gamma$ would be
an experimental challenge since $B\to X_s\gamma$ constitutes a
serious background. \cite{KSW} has suggested
several ways to overcome this problem, e.g. demanding a 
higher energy cut on $\gamma$, since $\gamma$ from 
$B\to X_d\gamma$ will be more energetic than that from 
$B\to X_s\gamma$.
Energy cuts can be used to separate $b\to s\gamma$
events from charmed background since there is a
high photon energy region that is inaccessible to
charmed states because of the mass of the charm quarks.
This method is not feasible for extracting $b\to d\gamma$
events from a $b\to s\gamma$ sample. Although the strange
quark mass is larger than the down quark mass, the respective lightest
hadronic single particle final states, $K^*$ and $\rho$, have almost 
the same mass (they actually overlap strongly). The lightest multi-particle
states are $K\pi$ and $\pi\pi$, respectively, but even here
effects such as bound state effects (neglected in \cite{KSW})
smear the spectra out over regions of the order of 200 MeV.
These effects constitute one of the major theoretical
uncertainties in the extraction of $BR(B\to X_s\gamma)$ from
the measured part of the spectrum, and they make a separation
of $b\to d\gamma$ and $b\to s\gamma$ via energy cuts impossible.
A comparison of the photon energy spectra for $b\to s\gamma$
and $b\to d\gamma$ was made in \cite{Ali92}, and showed that
the photon spectra for both decays are very similar.

A more promising approach constitute exclusive channels 
\cite{Alitalk,Ali}.
The improved $K/\pi$ separation at the $B$ factories may enable
the inclusive $B\to X_d\gamma$ decay to be reconstructed by summing 
over the relevant exclusive channels 
as done by CLEO in the measurement of $B\to X_s\gamma$. \cite{KSW} 
suggested using a semi-inclusive sample of 
$B \to \gamma + n\pi$ decays with a maximum of $n$ (say 5) mesons 
together with a corresponding measurement of
$B \to \gamma+K+ (n-1)\pi$. The ratio of the widths 
of the semi-inclusive samples would enable the
total inclusive rate to be deduced to a very good approximation.
Although the extraction of the branching ratio for $b\to d\gamma$
from exclusive channels might suffer additional uncertainties
with respect to $b\to s\gamma$ \cite{Donoghue:1997fv}
the asymmetry should not be affected by these.

If ${\cal A}^{d\gamma}_{CP}$ and ${\cal A}^{s\gamma}_{CP}$ 
cannot be separated, then only their sum can be measured.
In the context of the SM (with $m_s=m_d=0$)
the unitarity of the CKM matrix ensures that the sum
is zero \cite{neubert,soares}.
This relation holds only for the
short distance contribution, which is expected to
be dominant (c.f.\ introduction).
In the presence of new physics such a cancellation does not
occur, as will be
shown in section 4. As stressed earlier, a 
reliable prediction of 
${\cal A}^{d\gamma}_{CP}$ in a given model is necessary 
since it contributes to the measurement of ${\cal A}^{s\gamma}_{CP}$.
The CLEO result is sensitive to a weighted sum of CP asymmetries, 
given by:
\begin{equation}
{\cal A}^{exp}_{CP}=0.965{\cal A}^{s\gamma}_{CP}+0.02{\cal A}^{d\gamma}_{CP}
\label{CLEOeq} \end{equation}  
The latest measurement stands at $-27\% < {\cal A}^{exp}_{CP} < 10\%$ 
(90\% c.l.) \cite{CLEO}. 
The small coefficient of ${\cal A}^{d\gamma}_{CP}$ is caused by
the smaller BR$(B\to X_d\gamma)$ (assumed to be $1/20$ that
of BR$(B\to X_s\gamma)$) and inferior detection efficiencies, but 
may be partly compensated by the larger value for ${\cal A}^{d\gamma}_{CP}$.
We shall see that there can be both constructive and destructive
interference between the two terms in eq.~~(\ref{CLEOeq}). These effects will
be especially important for measurements in future high luminosity  
runs of $B$ factories, in which the precision is expected to reach
a magnitude where the $b\to d\gamma$ contribution becomes crucial.
For integrated luminosities of 200 fb$^{-1}$ (2500 fb$^{-1}$) \cite{alex}
anticipates a precision of $3\%(1\%)$ in the measurement of
${\cal A}^{exp}_{CP}$.

\section{Results}

We explore the effect of CP violating SUSY phases
on the direct CP asymmetry of the inclusive decay $B \to X_{d(s)} \gamma$.
We will show that the asymmetry ${\cal A}_{CP}^{d\gamma}$
may be quite different from the SM prediction
in a wide region of parameter space consistent with experimental
bounds from the Electric Dipole Moment (EDM) and BR($B \to X_{s}\gamma$).

In our analysis we adopt the ``effective SUSY'' model, proposed in
\cite{effSUSY}. 
It is instructive to consider the impact of unconstrained
SUSY phases on the inclusive decay $B\to X_d\gamma$ by
taking $A_t$ and $\mu$ complex. The same approach has been used in 
Ref.\cite{baek}, and it was shown that ${\cal A}^{s\gamma}_{CP}$ 
may lie in the range $-16\% \le {\cal A}^{s\gamma}_{CP}\le 16\%$.   

We vary the (SUSY) parameters in reasonable ranges and
respect the direct search lower limits on the masses of $\tilde t_1$,
$\chi^{\pm}$ by discarding generated points that do not
pass our cuts of $m_{\tilde t_1} > 90$ GeV and $m_{\chi^{\pm}_1} > 80$ GeV
in addition to the cut on $C_7$ mentioned in Sec.~3.
We vary $\rho$ and $\eta$ in the range allowed by present 
CKM fits for the SM \cite{CKM}. Note that in the effective SUSY
model one should strictly only include the constraint from
$|V_{ub}/V_{cb}|$, which corresponds to varying 
$\rho$ and $\eta$ in a semi-circular band in the 
$\rho-\eta$ plane. This enlarged parameter space has little effect
on our graphs, except for Fig.3, which will be commented on below.

\begin{figure}
\begin{center}
\caption{${\cal A}^{d\gamma}_{CP}$ against $m_{\tilde t_1}$}
\psfrag{XXX}{$m_{\tilde t_1}$}  \psfrag{YYY}
 {${\cal A}^{d\gamma}_{CP}$}
\includegraphics[width=9cm]{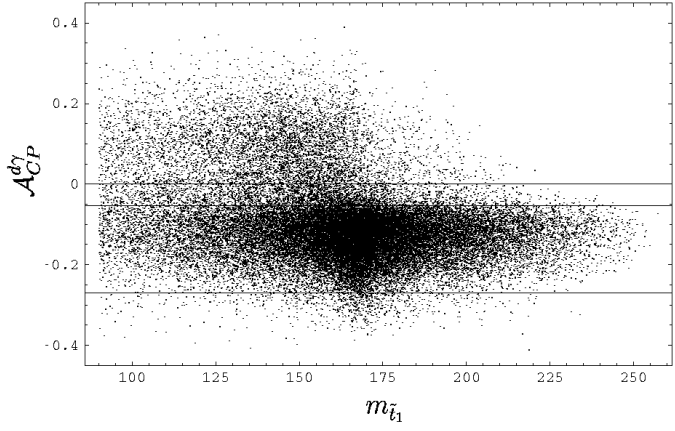}
\end{center}
\label{mst1}
\end{figure}

If the signal for the inclusive decay can be isolated
then a positive asymmetry would be a clear sign of new physics. 
In Fig.~\ref{mst1} we plot ${\cal A}^{d\gamma}_{CP}$ against 
$m_{\tilde t_1}$, which clearly shows that a light $\tilde t_1$ may 
drive ${\cal A}_{CP}^{d\gamma}$ positive, reaching maximal 
values close to $+40\%$. For $\tilde t_1$ heavier than 250 GeV
the ${\cal A}_{CP}^{d\gamma}$ lies within the SM range, which
is indicated by the two horizontal lines.

\begin{figure}
\begin{center}
\caption{${\cal A}^{d\gamma}_{CP}$ against $\tan\beta$}
\psfrag{XXX}{$\tan\beta$}  \psfrag{YYY}
 {${\cal A}^{d\gamma}_{CP}$}
\includegraphics[width=9cm]{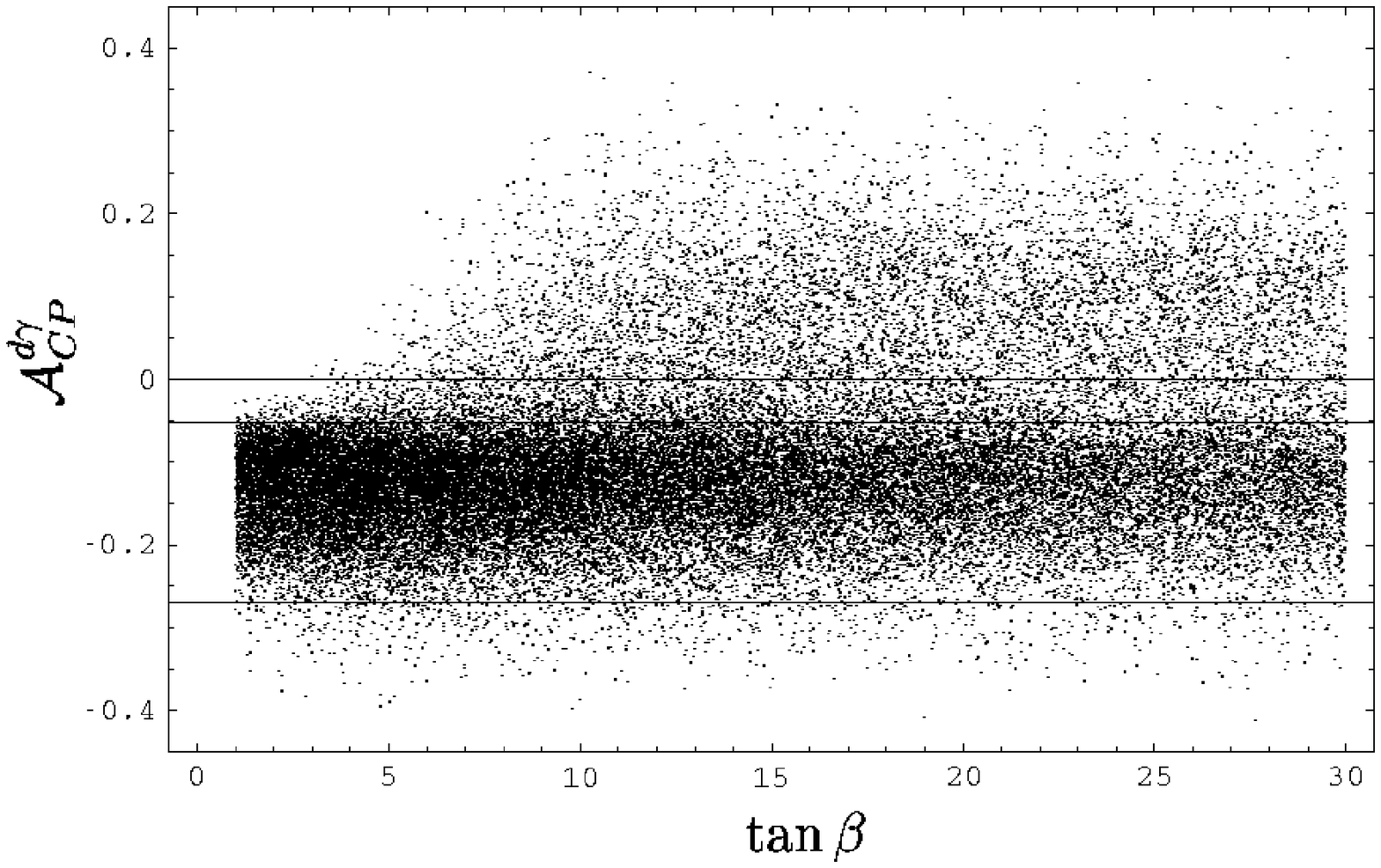}
\end{center}
\label{tbeta}
\end{figure}

We note that our upper limit of $+40\%$ is larger than 
the maximum value of $21\%$ attained in 
\cite{plb460}. The inclusion of the SUSY phases
has joined and expanded the two phenomenological regions found in 
\cite{plb460}, allowing CP asymmetries in the continuous region
$-40\%\le {\cal A}_{CP}^{d\gamma}\le 40\%$. 
In Fig.~\ref{tbeta} we show that the large positive asymmetries can be
found anywhere in the interval $5\le \tan\beta \le 30$, 
which is the region where the EDM constraint in \cite{PRL82} is 
comfortably satisfied. 

\begin{figure}
\begin{center}
\caption{${\cal A}^{d\gamma}_{CP}$ against ${\cal A}^{s\gamma}_{CP}$}
\psfrag{XXX}{${\cal A}^{s\gamma}_{CP}$}  \psfrag{YYY}
 {${\cal A}^{d\gamma}_{CP}$}
\includegraphics[width=9cm]{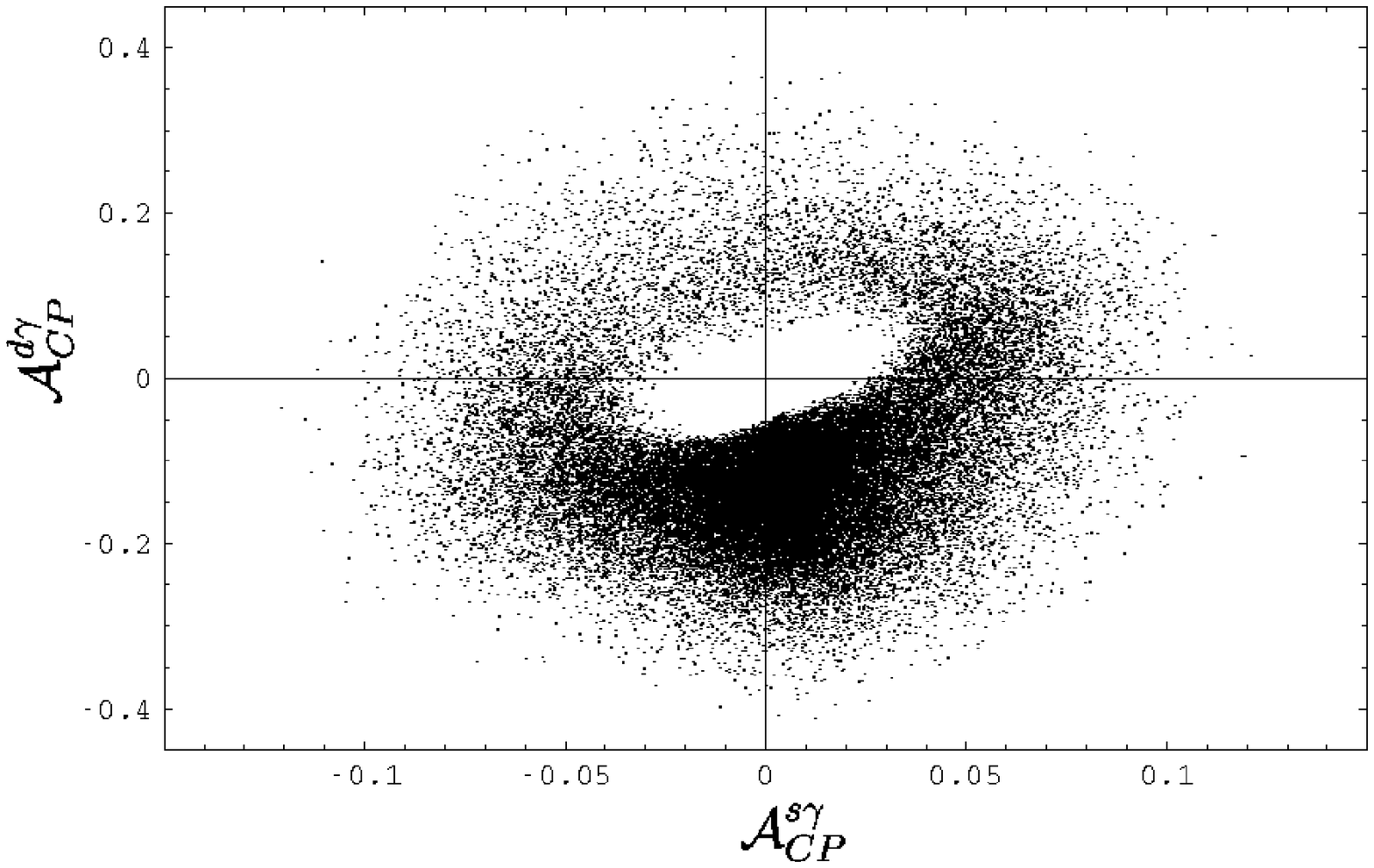}
\end{center}
\label{doughnut}
\end{figure}

If the signals from $b\to s\gamma$ and $b\to d\gamma$ cannot
be isolated then one must consider a combined signal.
In Fig.~\ref{doughnut} we plot ${\cal A}_{CP}^{d\gamma}$ against 
${\cal A}_{CP}^{s\gamma}$. The maximum values for
${\cal A}_{CP}^{s\gamma}$ agree with those found in \cite{baek}.
It can be seen that there is an inaccessible
region and the asymmetries can never simultaneously be
zero e.g. for ${\cal A}_{CP}^{d\gamma}\approx 0$, 
$|{\cal A}_{CP}^{s\gamma}|\ge 3\%$. This can be explained
from the fact that ${\cal A}_{CP}^{d\gamma}\approx 0$
would require $C_7$ to have a sizeable imaginary part in order to
cancel the large negative contribution from $\epsilon_d$.
The corresponding effect on ${\cal A}_{CP}^{s\gamma}$ would be to
cause a sizeable deviation from its small SM value.
Fig.~\ref{doughnut} shows that both ${\cal A}_{CP}^{s\gamma}$ and
${\cal A}_{CP}^{d\gamma}$ can have either sign, resulting
in constructive or destructive interference in eq.~(\ref{CLEOeq}).
If only the $|V_{ub}/V_{cb}|$ constraint is included in the CKM fits,
the enlarged parameter space for $\rho$ and $\eta$ allows much smaller
asymmetries for ${\cal A}_{CP}^{d,s\gamma}$. 
This is because smaller values of $\eta$ are now allowed, which reduces
the SM contribution to ${\cal A}_{CP}^{d,s\gamma}$. 
The choice of $\eta\to 0$ would correspond to points in the 
previously inaccessible region.

\begin{figure}
\begin{center}
\caption{$\Delta\Gamma_d+\Delta\Gamma_s$ against ${\rm Im} (C_7)$}
\psfrag{XXX}{${\rm Im}(C_7)$}  \psfrag{YYY}
 {$\Delta\Gamma_d+\Delta\Gamma_s$}
\includegraphics[width=9cm]{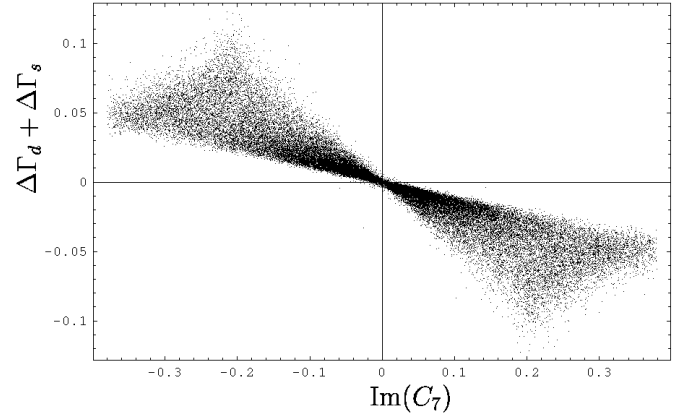}
\end{center}
\label{butterfly}
\end{figure}

In Fig.~\ref{butterfly} we plot $\Delta\Gamma_d+\Delta\Gamma_s$ 
(defined in eq.~(\ref{ACPdef})) against $\rm {Im} (C_7)$.
In the SM (as explained in Section 2) this sum would be exactly 
zero in the limit $m_s=m_d=0$ (neglecting the small
long distance contribution). From Fig.~\ref{butterfly} it can be seen 
that $\Delta\Gamma_d+\Delta\Gamma_s$ is close to 0 if $C_7$ is real, 
the slight deviation being caused by the imaginary parts of the
other Wilson coefficients. The effect of 
a non-zero $\rm{Im} (C_7)$ causes sizeable deviations from zero.

\begin{figure}
\begin{center}
\caption{Asymmetry measured by CLEO $({\cal A}^{exp}_{CP})$ 
 against ${\cal A}^{s\gamma}_{CP}$}
\psfrag{XXX}{${\cal A}^{s\gamma}_{CP}$}  \psfrag{YYY}
 {${\cal A}^{exp}_{CP}$}
\includegraphics[width=9cm]{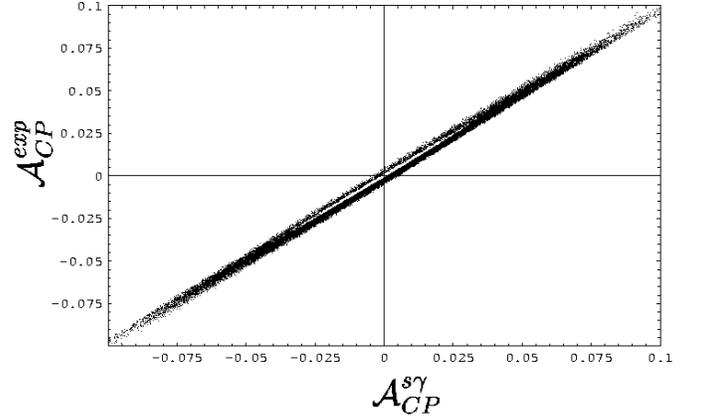}
\end{center}
\label{ACPCLEO}
\end{figure}

In Fig.~\ref{ACPCLEO} we plot the ${\cal A}_{CP}^{exp}$ 
(defined in eq.~(\ref{CLEOeq}))
against ${\cal A}_{CP}^{s\gamma}$. The right hand plot shows
a magnification of the area around the origin. The coefficient
of ${\cal A}_{CP}^{d\gamma}$ in eq.~(\ref{CLEOeq}) assumes that BR$(b\to d\gamma)$=
BR$(b\to s\gamma)/20$. Since this ratio of BRs is $\sim |V_{td}/V_{ts}|^2$,
which in turn is a function of the variables $\rho$ and $\eta$, we replace
the factor $1/20$ by the above ratio of CKM matrix elements.
If the contribution from ${\cal A}_{CP}^{d\gamma}$ were ignored
in eq.~(\ref{CLEOeq}), then Fig.~\ref{ACPCLEO} would be a straight 
line through the origin. The ${\cal A}_{CP}^{d\gamma}$ contribution 
broadens the line to a thin band of width $\approx 1\%$, an effect
which should be detectable at proposed higher luminosity runs of the
$B$ factories. 

Note that the width of the line is determined by the amount of
$b\to d\gamma$ admixture in the $b\to s\gamma$ sample,
eq.~(\ref{ACPdef}). In the case of the CLEO
measurement the admixture of $b\to d\gamma$ is about 2.5 times less than
the ``natural'' admixture (ratio of the branching ratios). If the
experimental analysis can be done with a natural admixture or even
a $b\to d\gamma$ enriched sample, the width of the line would 
be correspondingly
broader. Specifically, for the natural admixture the
line would be broadened by a factor of 2.5, making 
$b\to d\gamma$ a $2.5\%$ effect.
This effect is the same magnitude as the precision
attainable with an integrated luminosity of 200 fb$^{-1}$ at the
$B$ factories \cite{alex}. At this luminosity it will therefore be possible to
test the cancellation of the asymmetries as predicted by the SM.

\section*{Acknowledgments} 
We wish to thank the organizers for providing a very enjoyable
environment at a stimulating conference.

\newpage
\begin{figure}
\begin{center}
\caption{Asymmetry measured by CLEO $({\cal A}^{exp}_{CP})$ against 
 ${\cal A}^{s\gamma}_{CP}$}
\psfrag{XXX}{${\cal A}^{s\gamma}_{CP}$}  \psfrag{YYY}
 {${\cal A}^{exp}_{CP}$}
\includegraphics[width=9cm]{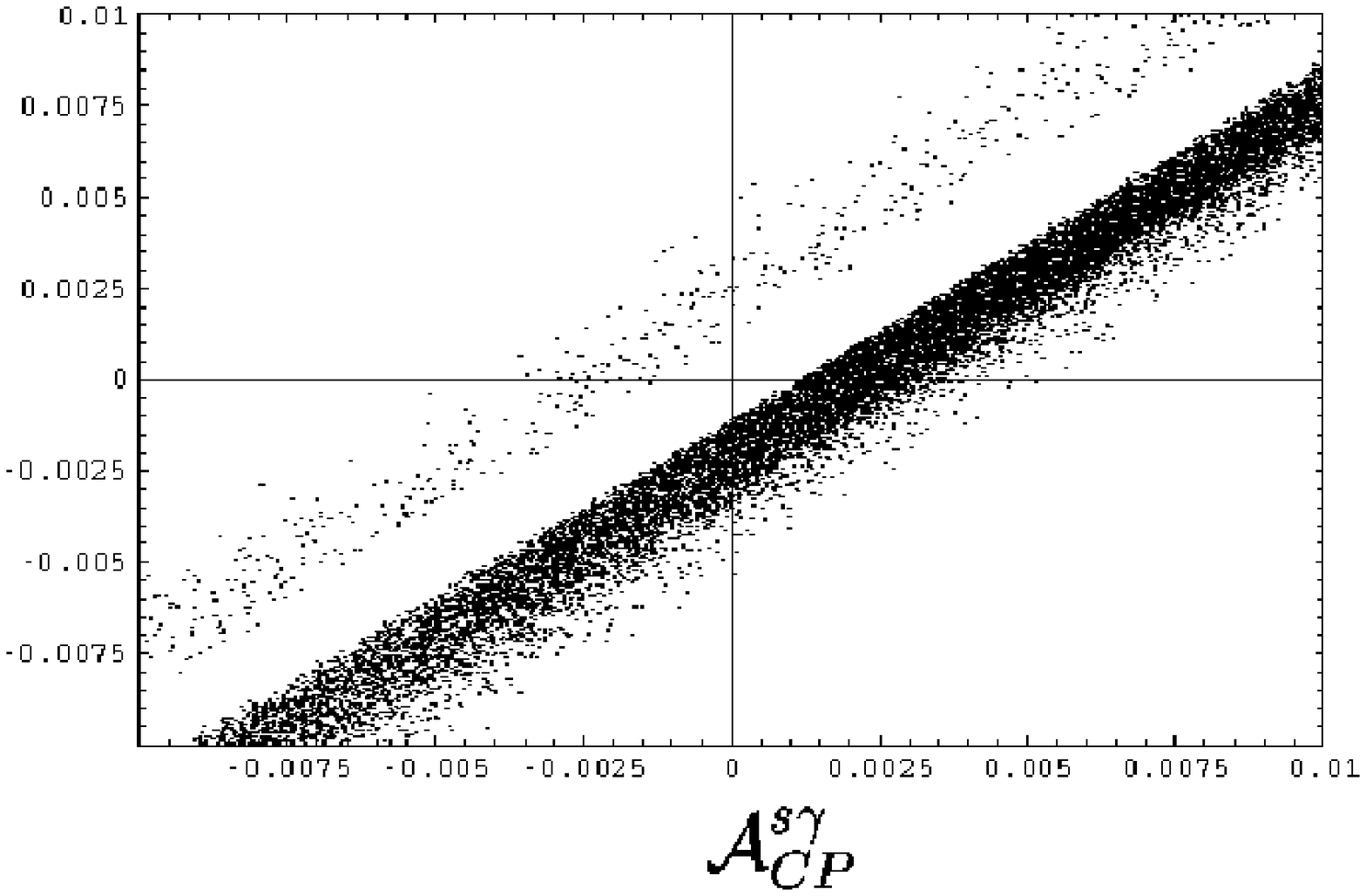}
\end{center}
\end{figure}

\end{document}